**Magnetoelectric susceptibility tensor of multiferroic TbMnO$_3$ with cycloidal antiferromagnetic structure in external field**

Igor V. Bychkov*, Dmitry A. Kuzmin and Sergey J. Lamekhov

*Chelyabinsk State University, 454001, Chelyabinsk, Br. Kashirinyh Street, 129, Russia*

Vladimir G. Shavrov

*The Institute of Radioengineering and Electronics of RAS, 125009, Moscow, Mokhovaya Street, 11-7, Russia*

Magnetoelectric, dielectric and magnetic susceptibility tensors of multiferroic TbMnO$_3$ with cycloidal antiferromagnetic structure in external electric and magnetic fields have been investigated with taking into account dynamics of spin, electro-dipole and acoustic subsystems. All components of tensors depend on values of external electric and magnetic fields. The possibility of control of electrodynamic properties of multiferroic TbMnO$_3$ with cycloidal antiferromagnetic structure by external electric and magnetic fields has been shown. The resonant interaction of spin, electro-dipole, electromagnetic and acoustic waves in such material is observed.

*E-mail address: bychkov@csu.ru

Nowadays multiferroics, materials with magnetic and electric ordering, attracts researchers' attention for their unusual physical properties[1-3]: existing of electromagnons and possibility to control it by an external electric field. Often multiferroics have a modulated magnetic structure, which contribute a number of features in the spectrum and dynamics of spin, acoustic and electromagnetic excitations in material[4-6]: band structure is observed, the nonreciprocity effect is manifested, i.e., difference between the velocity of wave transmission along and against the modulation axis. Orthorhombic multiferroic TbMnO$_3$ ($P_{bnm}$), for example, has a cycloidal antiferromagnetic structure at temperatures T < 28 K [7]. The coupled spin, electro-dipole and electromagnetic waves in TbMnO$_3$ with cycloidal antiferromagnetic structure had been theoretically investigated earlier[8,9], however the influence of external electric and magnetic fields on dynamic properties of such material are not studied enough. The present work is devoted to studying of electrodynamical properties of orthorhombic multiferroics placed in an external electric and magnetic fields of different directions. Derivations of dynamcis of the coupled excitations in the modulated magnetic structures are carried out in regime $L \gg a$, where $L = 2\pi/k$ is the period of modulated structure, $a$ is the lattice constant, $k$ is the wave number of modulated structure, when the phenomenological method is applicable.

Method of Lagrange has been used for investigating of dynamics of TbMnO$_3$ with cycloidal antiferromagnetic structure. The Lagrangian's expression is $L = E - F$, where $E$ - the kinetic energy, $F$ - the Ginzburg-Landau function. In case of orthorhombic multiferroic TbMnO$_3$:

$$E = \frac{1}{V}\int \frac{1}{2}\left(\mu \dot{\mathbf{A}}^2 + \tilde{\mu}\dot{\mathbf{M}} + \lambda \dot{\mathbf{P}}^2 + \rho \dot{\mathbf{u}}^2\right) d\mathbf{r}; \; F = \frac{1}{V}\int \{\frac{a}{2}\mathbf{A}^2 + \frac{w}{2}A_z^2 + \frac{u}{4}\mathbf{A}^4 + \frac{\gamma}{2}(\partial_y \mathbf{A})^2 + \frac{\alpha}{2}(\partial_y^2 \mathbf{A})^2 - $$
$$-\mathbf{MH} + \frac{\beta}{2}\mathbf{M}^2 + \frac{\lambda_1}{2}(\mathbf{AM})^2 + \frac{\lambda_2}{2}\mathbf{A}^2\mathbf{M}^2 + \frac{b}{2}\mathbf{P}^2 - \mathbf{PE} + \nu P_z\left(A_z \partial_y A_y - A_y \partial_y A_z\right) + $$
$$+c_{ijlm}u_{ij}u_{lm} + b_{ijlm}A_i A_j u_{lm} + d_{ijlm}P_i P_j u_{lm}\} d\mathbf{r}. \quad (1)$$

Where $\mathbf{M} = \mathbf{M_1} + \mathbf{M_2} + \mathbf{M_3} + \mathbf{M_4}$ is the magnetization of the crystal; $\mathbf{A} = \mathbf{M_1} - \mathbf{M_2} - \mathbf{M_3} + \mathbf{M_4}$ is the vector of antiferromagnetism; $\mathbf{P}$ is the vector of polarization; $u_{ij} = \left(\partial u_i / \partial x_j + \partial u_j / \partial x_i\right)/2$ is the tensor of deformations; $\mathbf{u}$ is the displacement vector; $w$ is the constants of anisotropy; $a, u, \beta, \lambda_1, \lambda_2$ is the constants of homogeneous exchange; $\alpha, \gamma$ is the constants of heterogeneous exchange; $b, \nu$ is the constants of electric and magnetoelectric interactions; $c_{ijlk}, b_{ijlk}, d_{ijlk}$ is the tensors of elasticity, magnetostriction and electrostriction; $\lambda = mv_c/z^2$, where $z$ and $m$ is the charge and the reduced mass of the elementary cell with the volume $v_c$; $\mu = \chi_\perp/8g^2M_0^2$, where $\chi_\perp$ is the static transversal magnetic susceptibility, $g$ is the gyromagnetic ratio, $M_0$ is the magnetization of sub lattices.

The ground state of TbMnO$_3$ with cycloidal antiferromagnetic structure is described by vectors of antiferromagnetism and polarization with following components: $P_{0x} = P_{0y} = 0$; $P_{0z} = P_0$; $A_{0x} = 0$; $A_{0y} = A_1 \sin ky$; $A_{0z} = A_2 \cos ky$. The coordinate axis's direction defined by $x\|a, y\|b, z\|c$. So, $y$-axis is the modulation axis, $z$-axis is the direction of the spontaneous polarization in TbMnO$_3$.

Expressions for determining the parameters of the ground state can be obtained from the minimum of energy. In case of external electric field $\mathbf{E}$ is collinear to a spontaneous polarization $\mathbf{P_{0z}}$ and external magnetic field have all non-zero components:

$$M_{0x}\left[\frac{\lambda_2}{2}\left(A_1^2+A_2^2\right)+\beta\right]=H_{0x};\; M_{0y,z}\left[\frac{\lambda_1}{2}A_{1,2}^2+\frac{\lambda_2}{2}\left(A_1^2+A_2^2\right)+\beta\right]=H_{0y,z};$$
$$\alpha k^3\left(A_1^2+A_2^2\right)+\frac{\gamma}{2}k\left(A_1^2+A_2^2\right)-\nu A_1 A_2 P_0 = 0;$$
$$p_3 P_0^3 + P_0\left(p_{1a1}A_1^2+p_{1a2}A_2^2+b\right)-\nu k A_1 A_2 = E_{0z}; \qquad (2)$$
$$a_{13}A_1^3+A_1\left(a_{11p}P_0^2+a_{11a2}A_2^2+a_{11my}\left(M_{0x}^2+M_{0y}^2+M_{0z}^2\right)+4\left(\alpha k^4+\gamma k^2+a\right)\right)-8\nu k P_0 A_2 = 0;$$
$$a_{23}A_2^3+A_2\left(a_{21p}P_0^2+a_{21a1}A_1^2+a_{21my}\left(M_{0x}^2+M_{0y}^2+M_{0z}^2\right)+4\left(\alpha k^4+\gamma k^2+a+w\right)\right)-8\nu k P_0 A_1 = 0;$$

Here, the following notation has been introduced: $\Delta = \det|c_{ij}|$, $p_3 = -8\Delta^{-1}(\mathbf{D_3}\mathbf{\Delta_{D3}})$, $p_{1a1} = -4\Delta^{-1}(\mathbf{B_2}\mathbf{\Delta_{D3}})$, $p_{1a2} = -4\Delta^{-1}(\mathbf{B_3}\mathbf{\Delta_{D3}})$, $a_{13} = 3u - 24\Delta^{-1}(\mathbf{B_2}\mathbf{\Delta_{B2}})$, $a_{11p} = -32\Delta^{-1}(\mathbf{D_3}\mathbf{\Delta_{B2}})$, $a_{11my} = 4(\lambda_1+\lambda_2)$, $a_{11a2} = a_{21a1} = u - B_{44}^2 c_{44}^{-1} - 8\Delta^{-1}(\mathbf{B_3}\mathbf{\Delta_{B2}})$, $a_{23} = 3u - 24\Delta^{-1}(\mathbf{B_3}\mathbf{\Delta_{B3}})$, $a_{21my} = 4\lambda_2$, $a_{21p} = -32\Delta^{-1}(\mathbf{D_3}\mathbf{\Delta_{B3}})$, $\mathbf{\Delta_{\Xi i}} = (\Delta_{\Xi i}^1, \Delta_{\Xi i}^2, \Delta_{\Xi i}^3)$, $\mathbf{\Xi_i} = (\Xi_{i1}, \Xi_{i2}, \Xi_{i3})$, $\Xi = B, D$, $i = 1, 2, 3$, $\Delta_{\Xi i}^j$ can be obtained from expression for $\Delta$ by changing of $j$ column to vector $\mathbf{\Xi_i}$.

The equilibrium deformations are heterogeneous:

$$u_{xx}^0 = -2\Delta^{-1}\left(\Delta_{B2}^1 A_1^2 \cos^2 ky + \Delta_{B3}^1 A_2^2 \sin^2 ky + \Delta_{D3}^1 P_0^2\right);$$
$$u_{yy}^0 = -2\Delta^{-1}\left(\Delta_{B2}^2 A_1^2 \cos^2 ky + \Delta_{B3}^2 A_2^2 \sin^2 ky + \Delta_{D3}^2 P_0^2\right);$$
$$u_{zz}^0 = -2\Delta^{-1}\left(\Delta_{B2}^3 A_1^2 \cos^2 ky + \Delta_{B3}^3 A_2^2 \sin^2 ky + \Delta_{D3}^3 P_0^2\right); \quad (3)$$
$$u_{xy}^0 = u_{xz}^0 = 0; \quad u_{yz}^0 = -B_{44}\left(2c_{44}\right)^{-1} A_1 A_2 \sin ky \cos ky.$$

For investigation of dynamic characteristics one have to take into account the system of Lagrange equations for **A**, **M**, **P** and **u**. Solving the system of equations by method of low oscillations, linearizing and using the form of harmonic series for variables, in approach of first harmonics for the waves, propagating along *y*-axis, the oscillating amplitudes of polarization **p** and antiferromagnetism **a** vectors can be expressed as $p_i = \alpha_{ij} e_j + \kappa_{ij}^{em} h_j$, $m_i = \chi_{ij} h_j + \kappa_{ij}^{me} e_j$, where, $\alpha_{ij}$, $\chi_{ij}$ is the tensors of dielectric and magnetic susceptibility, consequently, $\kappa_{ij}^{me} = \left(\kappa_{ij}^{em}\right)^*$ is the magnetoelectric susceptibility tensor (operation * means the complex conjugation).

The study shows that in case of $\mathbf{H} \parallel \mathbf{x}$, i.e. $\mathbf{H} = (H_{0x}, 0, 0)$, the tensors $\alpha_{ij}$ and $\chi_{ij}$ are diagonals with the following components:

$$\alpha_{xx} = \lambda^{-1}\left(\omega_{px}^2 - \omega^2\right)^{-1}; \quad \alpha_{zz} = \lambda^{-1}\Delta_{mx}\left(\omega^2 - \omega_l^2\right)\left(\Delta_{ay}^+ \Delta_{az}^- + \left[\Delta_{ay}^{+az}\right]^2\right)\left(\Delta_{mx}\Delta_{pz} - \Delta_{mx}^{pz}\Delta_{pz}^{mx}\right)^{-1};$$

$$\alpha_{yy} = \frac{\lambda^{-1}\left(\omega^2 - \omega_{tz}^2\right)\left(\Delta_{ay}^- \Delta_{az}^+ + \left[\Delta_{az}^{+ay}\right]^2\right)}{\Delta_{py}\left(\Delta_{ay}^- \Delta_{az}^+ + \left[\Delta_{az}^{+ay}\right]^2\right) + \omega_{D44}^z \omega_l^2 \omega_{tz}^2 \left[i\omega_{B44}^{zy}{}^2 \Delta_{az}^{+ay}\left(\omega_{D44}^z + \omega_{D32}^z\right) - \omega_{B44}^y{}^2 \omega_{D32}^z \Delta_{ay}^- - \omega_{B44}^z{}^2 \omega_{D44}^z \Delta_{az}^+\right]}; \quad (5)$$

$$\chi_{xx} = \tilde{\mu}^{-1}\lambda\alpha_{zz}\Delta_{pz}\Delta_{mx}^{-1}\left(\omega^2 - \omega_l^2\right)^{-1}; \quad \chi_{yy} = \tilde{\mu}^{-1}\Delta_{ax}^+ \left\{\left(\omega^2 - \omega_{my}^2\right)\Delta_{ax}^+ + \frac{1}{2}\omega_{1xy}^2\left(\omega^2 - \omega_{tx}^2\right)\right\}^{-1};$$

$$\chi_{zz} = \tilde{\mu}^{-1}\left(\omega^2 - \omega_{ax}^{-2}\right)\left\{\frac{1}{2}\omega_{1xz}^2 \omega_{2xz}^2 - \left(\omega^2 - \omega_{mz}^2\right)\left(\omega^2 - \omega_{ax}^{-2}\right)\right\}^{-1}$$

Here the following notation has been introduced: $\Delta_{ax}^+ = \omega_{B66}^y{}^2 \omega_{tx}^2 - \left(\omega^2 - \omega_{ax}^{+2}\right)\left(\omega^2 - \omega_{tx}^2\right)$,

$\Delta_{ay}^+ = \omega_{B22}^y{}^2 \omega_l^2 - \left(\omega^2 - \omega_{ay}^{+2}\right)\left(\omega^2 - \omega_l^2\right)$, $\Delta_{az}^- = \omega_{B32}^z{}^2 \omega_l^2 - \left(\omega^2 - \omega_{az}^{-2}\right)\left(\omega^2 - \omega_l^2\right)$, $\omega_{B66}^y{}^2 = \left(2\mu c_{66}\right)^{-1} B_{66}^2 A_1^2$,

$\left(\Delta_{az}^+, \Delta_{ay}^-\right) = \omega_{B44}^y{}^2 \omega_{tz}^2 - \left(\omega^2 - \left(\omega_{az}^{+2}, \omega_{ay}^{-2}\right)\right)\left(\omega^2 - \omega_{tz}^2\right)$, $\quad \Delta_{ay}^{+az} = i\Omega_-^2\left(\omega^2 - \omega_l^2\right) + i\omega_{B22}^y \omega_{B32}^z \omega_l^2$,

$\Delta_{az}^{+ay} = i\Omega_+^2\left(\omega^2 - \omega_{tz}^2\right) + i\omega_{B44}^z \omega_{B44}^y \omega_{tz}^2$, $\quad \omega_{B22}^y{}^2 = 2\left(\mu c_{22}\right)^{-1} B_{22}^2 A_1^2$, $\quad \omega_{B32}^z{}^2 = 2\left(\mu c_{22}\right)^{-1} B_{32}^2 A_2^2$,

$$\Delta_{py} = \omega_{D44}^{z}{}^{2}\omega_{tz}^{2} - \left(\omega^{2} - \omega_{py}^{2}\right)\left(\omega^{2} - \omega_{tz}^{2}\right), \quad \Delta_{pz} = \omega_{D32}^{z}{}^{2}\omega_{l}^{2} - \left(\omega^{2} - \omega_{pz}^{2}\right)\left(\omega^{2} - \omega_{l}^{2}\right), \quad \omega_{B44}^{y}{}^{2} = (2\mu c_{44})^{-1} B_{44}^{2} A_{1}^{2},$$

$$\Delta_{mx} = -\left(\omega^{2} - \omega_{l}^{2}\right)\left[\omega_{2xy}^{4}\Delta_{az}^{-} + \omega_{2xz}^{4}\Delta_{ay}^{+} + i2\omega_{2xy}^{2}\omega_{2xz}^{2}\Delta_{ay}^{+az}\right] - \left(\omega^{2} - \omega_{mx}^{2}\right)\left(\Delta_{ay}^{+}\Delta_{az}^{-} + \left[\Delta_{ay}^{+az}\right]^{2}\right),$$

$$\Delta_{mx}^{pz} = \sqrt{\lambda/\tilde{\mu}}\left\{\left(\omega^{2} - \omega_{l}^{2}\right)\left(\omega_{me}^{Ay2}\Delta_{ay}^{+az} + \omega_{me}^{Az2}\Delta_{az}^{-}\right)\left(\omega_{2xy}^{2} - i\omega_{2xz}^{2}\right) - i\omega_{D32}^{z}\omega_{l}^{2}\left(\omega_{B32}^{z}\Delta_{ay}^{+az} + \omega_{B22}^{y}\Delta_{az}^{-}\right)\left(\omega_{2xy}^{2} + i\omega_{2xz}^{2}\right)\right\},$$

$$\Delta_{pz}^{mx} = \sqrt{\tilde{\mu}/\lambda}\left(\omega^{2} - \omega_{l}^{2}\right)\left\{\begin{array}{l}\left(\omega^{2} - \omega_{l}^{2}\right)\left[\omega_{2xy}^{2}\left(\omega_{me}^{Az2}\Delta_{ay}^{+az} + i\omega_{me}^{Ay2}\Delta_{ay}^{+}\right) + \omega_{2xz}^{2}\left(\omega_{me}^{Ay2}\Delta_{ay}^{+az} - i\omega_{me}^{Az2}\Delta_{az}^{-}\right)\right] - \\ -\omega_{D32}^{z}\omega_{l}^{2}\left[\omega_{2xy}^{2}\left(\omega_{B22}^{y}\Delta_{ay}^{+az} - i\omega_{B32}^{z}\Delta_{ay}^{+}\right) - \omega_{2xz}^{2}\left(\omega_{B32}^{z}\Delta_{ay}^{+az} + i\omega_{B22}^{y}\Delta_{az}^{-}\right)\right]\end{array}\right\},$$

$$\omega_{B44}^{z}{}^{2} = (2\mu c_{44})^{-1} B_{44}^{2} A_{2}^{2}, \quad \omega_{B66}^{z}{}^{2} = (2\mu c_{66})^{-1} B_{66}^{2} A_{2}^{2}, \quad \omega_{B44}^{zy}{}^{2} = \omega_{B44}^{z}\omega_{B44}^{y}, \quad \omega_{D32}^{z}{}^{2} = 4(\lambda c_{22})^{-1} D_{32}^{2} P_{0}^{2},$$

$$\omega_{D44}^{z}{}^{2} = (\lambda c_{44})^{-1} D_{44}^{2} P_{0}^{2}, \quad \omega_{ij(y,z)}^{2} = (\mu\tilde{\mu})^{-1/2} \lambda_{i} M_{0j} A_{(1,2)}, \quad i = 1, 2, \, j = x, y, z, \quad \omega_{me}^{A(y,z)2} = 2\mu^{-1}\nu A_{(1,2)} k$$

$$\Omega_{\pm}^{2} = \mu^{-1}\left\{\left[u - B_{44}^{2}(2c_{44})^{-1}\right] A_{1} A_{2} \pm 2\nu P_{0} k\right\}, \quad \omega_{tx} = s_{tx} q, \quad \omega_{tz} = s_{tz} q, \quad \omega_{l} = s_{l} q, \quad s_{tx} = \sqrt{2c_{66}/\rho}, \quad s_{tz} = \sqrt{2c_{44}/\rho},$$

$s_{l} = \sqrt{c_{22}/\rho}$ are the velocities of transversal $x$ and $y$ polarized and longitudinal acoustic waves, $q$ is the wave number of the propagating wave, $\omega_{pi}^{2} = \lambda^{-1}\left(b + 4\left(\mathbf{D_{i}}, \mathbf{u^{0p}}\right)\right), i = x, y, z,$

$$\mathbf{u^{0p}} = -2\Delta^{-1}\left(\Delta_{D3}^{1}, \Delta_{D3}^{2}, \Delta_{D3}^{3}\right), \quad \omega_{mx}^{2} = \tilde{\mu}^{-1}\left[\beta + \lambda_{2}\left(A_{1}^{2} + A_{2}^{2}\right)/2\right], \quad \omega_{my}^{2} = \tilde{\mu}^{-1}\left[\beta + \lambda_{2}\left(A_{1}^{2} + A_{2}^{2}\right)/2 + \lambda_{1} A_{1}^{2}/2\right],$$

$$\omega_{a(x,y,z)}^{\pm}{}^{2} = \mu^{-1}\left[\begin{array}{l}a + \lambda_{1} M_{0(x,y,z)} + \lambda_{2} \mathbf{M}^{2} + 4\left(\mathbf{B}_{(1,2,3)}, \mathbf{u^{0P}}\right) + \gamma k^{2} + \alpha k^{4} + \\ +\left(u\{A_{1}^{2} + A_{2}^{2}\} + 4\left(\mathbf{B}_{(1,2,3)}, \mathbf{u^{0A1}} + \mathbf{u^{0A2}}\right)\right)/2 \pm \left(u\{A_{1}^{2} - A_{2}^{2}\} + 4\left(\mathbf{B}_{(1,2,3)}, \mathbf{u^{0A1}} - \mathbf{u^{0A2}}\right)\right)/4\end{array}\right],$$

$$\omega_{mz}^{2} = \tilde{\mu}^{-1}\left[\beta + \lambda_{2}\left(A_{1}^{2} + A_{2}^{2}\right)/2 + \lambda_{1} A_{2}^{2}/2\right].$$

The magnetoelectric susceptibility tensor $\kappa_{ij}^{me}$ has only one non-zero component:

$$\kappa_{xz}^{me} = -\alpha_{zz}\Delta_{mx}^{pz}\Delta_{mx}^{-1} \tag{6}$$

In case of $\mathbf{H} \parallel \mathbf{y}$, i.e. $\mathbf{H} = (0, H_{0y}, 0)$, the tensors $\alpha_{ij}$ and $\chi_{ij}$ also are diagonals with the components:

$$\alpha_{xx} = \lambda^{-1}\left(\omega_{px}^{2} - \omega^{2}\right)^{-1}; \quad \alpha_{yy} = \lambda^{-1}\Delta_{mz}\left(\omega^{2} - \omega_{tz}^{2}\right)\left(\Delta_{ay}^{-}\Delta_{az}^{+} + \left[\Delta_{az}^{+ay}\right]^{2}\right)\left(\Delta_{mz}\tilde{\Delta}_{py} - \Delta_{mz}^{py}\Delta_{py}^{mz}\right)^{-1};$$

$$\alpha_{zz} = \frac{\Delta_{my}\left(\omega^{2} - \omega_{l}^{2}\right)\left(\Delta_{ay}^{+}\Delta_{az}^{-} + \left[\Delta_{ay}^{+az}\right]^{2}\right)}{\lambda\left(\Delta_{my}\tilde{\Delta}_{pz} - \Delta_{my}^{pz}\Delta_{pz}^{my}\right)}; \quad \chi_{xx} = \Delta_{ax}^{+}\tilde{\mu}^{-1}\left\{\left(\omega_{mx}^{2} - \omega^{2}\right)\Delta_{ax}^{+} + \frac{1}{2}\omega_{1yy}^{4}\left(\omega_{tx}^{2} - \omega^{2}\right)\right\}^{-1}; \tag{7}$$

$$\chi_{yy} = \lambda\alpha_{zz}\Delta_{pz}\tilde{\mu}^{-1}\Delta_{my}^{-1}\left(\omega^{2} - \omega_{l}^{2}\right)^{-1}; \chi_{zz} = \lambda\alpha_{yy}\Delta_{py}\tilde{\mu}^{-1}\Delta_{mz}^{-1}\left(\omega^{2} - \omega_{tz}^{2}\right)^{-1}.$$

Note, that in (7):

$$\tilde{\Delta}_{py} = \Delta_{py}\left(\Delta_{ay}^-\Delta_{az}^+ + \left[\Delta_{az}^{+ay}\right]^2\right) - \omega_{D44}^{z}{}^2\omega_{tz}^4\left[2i\omega_{B44}^{zy}{}^2\Delta_{az}^{+ay} + \omega_{B44}^{y}{}^2\Delta_{ay}^- + \omega_{B44}^{z}{}^2\Delta_{az}^+\right], \quad \tilde{\omega}_{me}^{Ay2} = \omega_{me}^{Ay2} - \omega_{B32}^{z}\omega_{D32}^{z}$$

$$\tilde{\Delta}_{pz} = \Delta_{pz}\left(\Delta_{ay}^+\Delta_{az}^- + \left[\Delta_{ay}^{+az}\right]^2\right) +$$
$$+ \left\{\begin{array}{l} i\tilde{\omega}_{me}^{Ay2}\left[\Delta_{ay}^+\left(\omega_{me}^{Ay2}\left(\omega^2-\omega_l^2\right)+\omega_{B32}^{z}\omega_{D32}^{z}\omega_l^2\right) - \Delta_{ay}^{+az}\left(\omega_{me}^{Az2}\left(\omega^2-\omega_l^2\right)-\omega_{B22}^{y}\omega_{D32}^{z}\omega_l^2\right)\right] - \\ -\tilde{\omega}_{me}^{Az2}\left[\Delta_{az}^-\left(\omega_{me}^{Az2}\left(\omega^2-\omega_l^2\right)-\omega_{B22}^{y}\omega_{D32}^{z}\omega_l^2\right)-\Delta_{ay}^{+az}\left(\omega_{me}^{Ay2}\left(\omega^2-\omega_l^2\right)+\omega_{B32}^{z}\omega_{D32}^{z}\omega_l^2\right)\right]\end{array}\right\},$$

$$\Delta_{my} = -2\left(\omega^2-\omega_l^2\right)\left[\left(\omega_{1yy}^2+\omega_{2yy}^2\right)^2\Delta_{az}^- + \omega_{2yz}^4\Delta_{ay}^+ + 2i\left(\omega_{1yy}^2+\omega_{2yy}^2\right)\omega_{2yz}^2\Delta_{ay}^{+az}\right] -$$
$$-\left(\omega^2-\omega_{my}^2\right)\left(\Delta_{ay}^+\Delta_{az}^- + \left[\Delta_{ay}^{+az}\right]^2\right), \quad \tilde{\omega}_{me}^{Az2} = \omega_{me}^{Az2} - \omega_{B22}^{y}\omega_{D32}^{z}$$

$$\Delta_{mz} = 2^{-1}\left(\omega^2-\omega_{tz}^2\right)\left[\omega_{1yy}^4\Delta_{ay}^- + \omega_{1yz}^4\Delta_{az}^+\right] - \left(\omega^2-\omega_{mz}^2\right)\left(\Delta_{ay}^-\Delta_{az}^+ + \left[\Delta_{ay}^{+az}\right]^2\right),$$

$$\Delta_{my}^{pz} = \sqrt{\lambda/\tilde{\mu}}\left\{\begin{array}{l}\left[\omega_{me}^{Ay2}\left(\omega^2-\omega_l^2\right)+\omega_{B32}^{z}\omega_{D32}^{z}\omega_l^2\right]\left[\Delta_{ay}^{+az}\left(\omega_{1yy}^2+\omega_{2yy}^2\right)-i\omega_{2yz}^2\Delta_{ay}^+\right]+\\ +\left[\omega_{me}^{Az2}\left(\omega^2-\omega_l^2\right)-\omega_{B22}^{y}\omega_{D32}^{z}\omega_l^2\right]\left[\Delta_{az}^-\left(\omega_{1yy}^2+\omega_{2yy}^2\right)-i\omega_{2yz}^2\Delta_{ay}^{+az}\right]\end{array}\right\},$$

$$\Delta_{pz}^{my} = 2\sqrt{\tilde{\mu}/\lambda}\left(\omega^2-\omega_l^2\right)\left[\left(\omega_{1yy}^2+\omega_{2yy}^2\right)\tilde{\omega}_{me}^{Az2}\Delta_{az}^- + \omega_{2yz}^2\tilde{\omega}_{me}^{Ay2}\Delta_{ay}^+ + i\left\{\tilde{\omega}_{me}^{Ay2}\left(\omega_{1yy}^2+\omega_{2yy}^2\right)+\omega_{2yz}^2\tilde{\omega}_{me}^{Az2}\right\}\Delta_{ay}^{+az}\right],$$

$$\Delta_{mz}^{py} = 2^{-1}\sqrt{\lambda/\tilde{\mu}}\omega_{D44}^{z}\omega_{tz}^2\left[i\omega_{1yz}^2\omega_{B44}^{z}\Delta_{az}^+ - \omega_{1yy}^2\omega_{B44}^{y}\Delta_{ay}^- - \Delta_{ay}^{+az}\left(\omega_{1yz}^2\omega_{B44}^{z} + i\omega_{1yy}^2\omega_{B44}^{y}\right)\right],$$

$$\Delta_{pz}^{my} = \omega_{D44}^{z}{}^2\omega_{tz}^4\sqrt{\tilde{\mu}/\lambda}\left[\omega_{B44}^{y}{}^2\Delta_{az}^- - \omega_{B44}^{z}{}^2\Delta_{ay}^+ - 2i\omega_{B44}^{y}\omega_{B44}^{z}\Delta_{ay}^{+az}\right].$$ The other symbols have been introduced above.

In this case there are two non-zeros components of magnetoelectric susceptibility tensor:

$$\kappa_{yz}^{me} = -\alpha_{zz}\Delta_{my}^{pz}\Delta_{my}^{-1}; \quad \kappa_{zy}^{me} = -\alpha_{yy}\Delta_{mz}^{py}\Delta_{mz}^{-1}; \tag{8}$$

In case of $\mathbf{H} \parallel \mathbf{z}$, i.e. $\mathbf{H} = (0, 0, H_{0z})$, the tensors $\alpha_{ij}$ and $\chi_{ij}$ are:

$$\alpha_{xx} = \lambda^{-1}\left(\omega_{px}^2-\omega^2\right)^{-1}; \quad \alpha_{yy} = \lambda^{-1}\Delta_{mz}\left(\omega^2-\omega_{tz}^2\right)\left(\Delta_{ay}^-\Delta_{az}^+ + \left[\Delta_{az}^{+ay}\right]^2\right)\left(\Delta_{my}\tilde{\Delta}_{py} - \Delta_{my}^{py}\Delta_{py}^{my}\right)^{-1};$$

$$\alpha_{zz} = \frac{\Delta_{my}\left(\omega^2-\omega_l^2\right)\left(\Delta_{ay}^+\Delta_{az}^- + \left[\Delta_{ay}^{+az}\right]^2\right)}{\lambda\left(\Delta_{mz}\tilde{\Delta}_{pz} - \Delta_{mz}^{pz}\Delta_{pz}^{mz}\right)}; \quad \chi_{xx} = \tilde{\mu}^{-1}\left(\omega^2-\omega_{ax}^{-2}\right)\left\{\frac{1}{2}\omega_{1zz}^4 - \left(\omega^2-\omega_{mx}^2\right)\left(\omega^2-\omega_{ax}^{-2}\right)\right\}^{-1}; \quad (9)$$

$$\chi_{yy} = \lambda\alpha_{yy}\Delta_{py}\tilde{\mu}^{-1}\Delta_{mz}^{-1}\left(\omega^2-\omega_{tz}^2\right)^{-1}; \quad \chi_{zz} = \alpha_{zz}\Delta_{pz}\tilde{\mu}^{-1}\Delta_{my}^{-1}\left(\omega^2-\omega_l^2\right)^{-1};$$

In (9) the following notation has been used:

$$\Delta_{my} = -2^{-1}\left(\omega^2 - \omega_{tz}^2\right)\left[\omega_{1zy}^4 \Delta_{az}^- + 2\omega_{1zy}^2 \omega_{1zz}^2 \Delta_{ay}^+ + i\left(\omega_{1zz}^2 + \omega_{1zy}^2\right)\omega_{zy}^2 \Delta_{ay}^{+az}\right] - \left(\omega^2 - \omega_{my}^2\right)\left(\Delta_{ay}^+ \Delta_{az}^- + \left[\Delta_{ay}^{+az}\right]^2\right),$$

$$\Delta_{mz} = -2\left(\omega^2 - \omega_l^2\right)\left[\omega_{2zy}^4 \Delta_{az}^- + \left(\omega_{1zz}^2 + \omega_{2zz}^2\right)^2 \Delta_{ay}^+ + 2i\left(\omega_{1zz}^2 + \omega_{2zz}^2\right)\omega_{2zy}^2 \Delta_{ay}^{+az}\right] -$$
$$- \left(\omega^2 - \omega_{mz}^2\right)\left(\Delta_{ay}^+ \Delta_{az}^- + \left[\Delta_{ay}^{+az}\right]^2\right),$$

$$\Delta_{my}^{py} = -2^{-1}\omega_{D44}^z \omega_{tz}^2 \sqrt{\lambda/\tilde{\mu}}\left[\omega_{1zz}^2 \omega_{B44}^z \Delta_{az}^+ + \omega_{1zy}^2 \omega_{B44}^y \Delta_{ay}^- + i\Delta_{ay}^{+az}\left(\omega_{1zz}^2 \omega_{B44}^z + \omega_{1zy}^2 \omega_{B44}^y\right)\right],$$

$$\Delta_{py}^{my} = -\omega_{D44}^z \omega_{tz}^2 \omega_{1zy}^2 \sqrt{\tilde{\mu}/\lambda}\left(\omega^2 - \omega_{tz}^2\right)\left[2\omega_{B44}^z \Delta_{az}^+ + \omega_{B44}^y \Delta_{ay}^- + i\Delta_{ay}^{+az}\left(\omega_{B44}^z + 2\omega_{B44}^y\right)\right],$$

$$\Delta_{mz}^{pz} = \sqrt{\lambda/\tilde{\mu}}\left\{\begin{array}{l}\left[\omega_{me}^{Ay2}\left(\omega^2 - \omega_l^2\right) + \omega_{B32}^z \omega_{D32}^z \omega_l^2\right]\left[\omega_{2zy}^2 \Delta_{ay}^{+az} + i\left(\omega_{1zz}^2 + \omega_{2zz}^2\right)\Delta_{ay}^+\right] + \\ + \left[\omega_{me}^{Az2}\left(\omega^2 - \omega_l^2\right) - \omega_{B22}^y \omega_{D32}^z \omega_l^2\right]\left[\omega_{2zy}^2 \Delta_{az}^- - i\left(\omega_{1zz}^2 + \omega_{2zz}^2\right)\Delta_{ay}^{+az}\right]\end{array}\right\},$$

$$\Delta_{pz}^{mz} = 2\sqrt{\tilde{\mu}/\lambda}\left(\omega^2 - \omega_l^2\right)\left\{\begin{array}{l}i\left[\omega_{me}^{Ay2}\left(\omega^2 - \omega_l^2\right) + \omega_{B32}^z \omega_{D32}^z \omega_l^2\right]\left[\omega_{2zy}^2 \Delta_{ay}^{+az} + i\left(\omega_{1zz}^2 + \omega_{2zz}^2\right)\Delta_{ay}^+\right] + \\ + \left[\omega_{me}^{Az2}\left(\omega^2 - \omega_l^2\right) - \omega_{B22}^y \omega_{D32}^z \omega_l^2\right]\left[\omega_{2zy}^2 \Delta_{az}^- + i\left(\omega_{1zz}^2 + \omega_{2zz}^2\right)\Delta_{ay}^{+az}\right]\end{array}\right\}.$$

Also there are two non-zeros components of tensor $\kappa_{ij}^{me}$:

$$\kappa_{yy}^{me} = -\alpha_{yy}\Delta_{my}^{py}\Delta_{mz}^{-1}; \quad \kappa_{zz}^{me} = -\alpha_{zz}\Delta_{mz}^{pz}\Delta_{my}^{-1}. \tag{10}$$

Note, that without an external magnetic field the tensor of magnetoelectric susceptibility has no non-zeros components. So, an external magnetic field activates some components of the tensor depending on the direction of field.

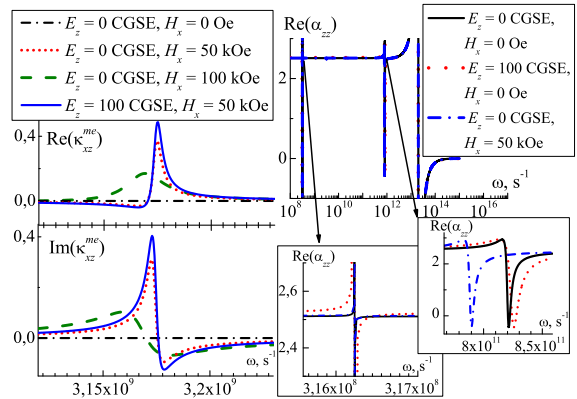

FIG. 1. Frequency dependence of some components of the susceptibility tensor.

(10 kOe = 1 T, 100 CGSE ≈ 30 kV/cm)

One can see that all components of all tensors have a resonant type. This effect is manifested in cause of resonant interaction of subsystems of TbMnO$_3$. As well as $k$, $A_1$, $A_2$, $P_0$ and $M_{0i}$ are the functions from $\mathbf{H}$ and $E_z$, defined from solving of the system of equations (2), the resonant frequency values depend on values of both electric and magnetic fields and its direction. Fig. 1 shows this dependence. Let us now estimate some dependences. From (2), $A_1^2 \approx A_{10}^2 - (u\beta^2)^{-1}\lambda_2\mathbf{H}^2$, $A_2^2 \approx A_{20}^2 - (u\beta^2)^{-1}(\lambda_1 - 2\lambda_2)\mathbf{H}^2$, $M_{0i} \approx \beta^{-1}H_i$  $P_0 \approx k\nu b^{-1}A_1A_2 + b^{-1}E_z$, $A_{10}^2 = (2u)^{-1}(w - 2L_1)$, $A_{20}^2 = -(2u)^{-1}(w + 2L_1)$, $k^2 \approx -\gamma/2\alpha$, $L_1 = a - \gamma^2/4\alpha < 0$. For values of constants[9-12] $\gamma \sim -10^{-14}$ cm$^2$, $\alpha \sim 10^{-28}$ cm$^4$, $a \sim -100$, $u \sim 0.1$, $w \sim 10$, $\beta \sim 100$, $b \sim 0.4$, $\nu \sim 10^{-9}$, $\lambda_{1,2} \sim 10^{-4}$ we have $A_{1,20}^2 \sim 10^3$ Oe$^2$, $k \sim 10^7$ cm$^{-1}$. So, for example, $\omega_{1xz}^2 \approx (\mu\tilde{\mu})^{-1/2}\beta^{-1}\lambda_1 H_x \left(A_{20}^2 - (u\beta^2)^{-1}(\lambda_1 - 2\lambda_2)H_x^2\right)^{1/2}$ vary in range of $0 \leq \omega_{1xz}^2 \leq (\omega_{1xz}^2)_{max}$ with changing of $H_x$, and take a maximum $(\omega_{1xz}^2)_{max} \sim 10^{21}$ s$^{-2}$ in field $H_x \sim 10^5$ Oe. Decreasing of field to $H_x \sim 10^4$ Oe will decrease this value to $\omega_{1xz}^2 \sim 10^{19}$ s$^{-2}$. The frequency $\Omega_\pm$ changes with the electric field on value about $\Delta\Omega \approx (\nu k\mu^{-1}b^{-1}E_z)^{1/2} \sim 10^{10}$ s$^{-1}$ for $E_z \sim 100$ CGSE (about 30 kV/cm). It shifts the resonant frequency corresponding to interaction of electromagnetic wave with antiferromagnetic subsystem to lower frequencies on value about $\omega_{1xz}$ and to higher frequencies on $\Delta\Omega$. The dielectric and magnetic susceptibility tensors have different expressions for their components in magnetic field of different directions. As well as reflectance and transmittance of electromagnetic waves depends on the dielectric and magnetic susceptibility, these characteristics also depend on values of electric and magnetic fields and its direction. So, it is possible to control the electrodynamic properties of TbMnO$_3$ by external electric and magnetic fields.